\documentclass[12pt]{aastex}

\begin{document}

\title{Calculation of optimized apodizers for a Terrestrial
Planet Finder coronagraphic telescope} 

\author{R. Gonsalves} 

\affil{Department of Electrical Engineering and Computer Science \\
Tufts University, Medford, MA 02155 \\
bobg@eecs.tufts.edu}

\author{P. Nisenson} 

\affil{Harvard-Smithsonian Center for Astrophysics \\
 60 Garden St., Cambridge, MA 01803 \\
pnisenson@cfa.harvard.edu}

\begin{abstract}

One of two approaches to implementing NASA's Terrestrial Planet Finder
is to build a space telescope that utilizes the techniques of
coronagraphy and apodization to suppress diffraction and image
exo-planets.  We present a method for calculation of a telescope's
apodizer which suppresses the side lobes of the image of a star so as
to optimally detect an Earth-like planet.  Given the shape of a
telescope's aperture and given a search region for a detector, we
solve an integral equation to determine an amplitude modulation (an
apodizer) which suppresses the star's energy in the focal plane search
region.  The method is quite general and yields as special cases the
product apodizer reported by Nisenson and Papaliolios (2001) and the
Prolate spheroidal apodizer of Kasdin et al (2002), and Aime et al
(2002).  We show computer simulations of the apodizers and the
corresponding point spread functions for various aperture-detector
configurations.

\end{abstract}


\section{Introduction}

A central problem in the search for Earth-like planets is to suppress
the light from a star so as to reveal a planet which might be 10 to 12
orders of magnitude fainter than the star and might be in quite close
to the star. In earlier papers (Nisenson and Papliolios, 2001;
Nisenson et al, 2003) we showed that apodizing a telescope pupil
reduced the diffraction of a telescope that would allow exo-planet
detection. In addition to diffraction suppression, extremely low
scattering optical surfaces at frequencies of 3 to 30 cycles/mirror
are necessary. Obtaining such low scattering will require a combination
of super-polishing and adaptive optic correction.  In this paper, we 
describe a technique  for designing an apodizer that minimizes 
diffraction in a chosen region of the focal plane while maximizing 
the throughput of the apodizer.

The image of a star (unaberrated by the atmosphere) is the point
spread function (PSF) of the telescope.  We want to design a real,
non-negative, continuous tone mask, i.e., an apodizer, to place in the
aperture (or a relay plane of the aperture) such that the PSF has a
compact, central peak with minimal spread beyond the core of the PSF.

This problem was first studied by Slepian \& Pollack (1961) in the context of
radar, communications, and superresolution.  The authors show that the
band-limited signal, $\phi(t)$, which has the most energy in time T
satisfies the integral equation:

\begin{equation}
  	\lambda\phi(t) = \int  \phi(s) w(s) a(t-s) ds
\end{equation}

where $w(t)$ is a zero-one function which time-limits $\phi(t)$ to T
seconds and the Fourier transform, $A(f)$, of $a(t)$ is another
zero-one function which band-limits $\phi(t)$.  The solutions
(eigenfunctions) of this integral equation are prolate spheroidal (PS)
functions and $\phi(t)$ is the eigenfunction with the largest
eigenvalue.  The authors show that $\phi(t)$, inside T, is identical
to its Fourier Transform, $\Phi(f)$, suitably scaled in f.  Kaiser
(1966) presented an approximation to $\phi(t)$ and subsequent authors
(Oppenheim and Schafer, 1989; Aime et al, 2002) have used the Kaiser
window because $\phi(t)$ is difficult to compute.

Papoulis (1968) studied the same problem in the optical context and
showed that $\Phi(u)$, the Fourier transform of $\phi(t)$, is the
optimal one-dimensional apodizer.  He showed that the optimal apodizer
for a square aperture is the product $\Phi(u) \Phi(v)$, which is the
form proposed by Nisenson and Papaliolios (2001).  He also shows that
the optimal apodizer for a circularly symmetric aperture is a sum of
the first and second order Bessel functions.  Other authors (Kasdin et
al, 2002 and Aime et al, 2001), have shown the optimality of a PS
apodizer.  Kasdin et al (2002) suggest a PS-shaped, binary aperture
with the key advantage that the binary aperture may be easier to
manufacture.

In this paper we present an iterative calculation of an optimal
apodizer.  The method does not require the Kaiser approximation and it
can accommodate any special features of the measurement process, such
as the likely position of a planet in reference to its star.  We use a
very simple iterative procedure to calculate the apodizer, an
iterative algorithm that has its roots in linear system theory and is
closely related to the well-known Gerchberg-Saxton algorithm (Gerchberg and
Saxton, 1972).

\section{The Algorithm}

We use the method of alternating projections which was introduced by
von Neumann (1950).  We write Equation (1) in the compact form

\begin{equation}	
  \lambda \phi =  R \phi  ,	
\end{equation}

where R is an operator, sometimes called the kernel.  R space-limits
the two-dimensional function $\phi(x,y)$ with a specified spatial
mask, $w(x,y)$, then band-limits it with a specified spatial frequency
mask, $A(u,v)$.  The latter is determined by the shape of the aperture.

To solve Eq. (2) we repeatedly impose R on an initial $\phi(x,y)$
until Eq. (2) is satisfied.  More specifically, we calculate

\begin{equation}	
	f(x,y) = \phi(x,y) w(x,y),			
\end{equation}

then we calculate the Fourier transform of f(x,y), namely F(u,v), and
calculate

\begin{equation}
	\Phi(u,v) = F(u,v) A(u,v) . 		
\end{equation}

The inverse Fourier transform of $\Phi(u,v)$ is $\phi(x,y)$, which
completes one iteration of the algorithm.

We rely on the fact that n iterations will produce an expansion of
$\phi(x,y)$ in eigenfunctions of Eq. 2 with coefficients which are
multiplied by eigenvalues raised to the power n.  Eventually, only the
eigenfunction with the largest eigenvalue will emerge, a property
which is well-known.  The algorithm is described by the figures and
text in Papoulis' 1968 book.

Because the operation R uses a forward Fourier transform, a masking
operation, an inverse Fourier transform, and another masking
operation, the algorithm is similar to the iterative transform
algorithm generally known as the Gerchberg-Saxton algorithm (Gerchberg
and Saxton, 1972).

We emphasize that the functions $w(x,y)$ and $A(u,v)$ can be any
desired shape, which gives us the opportunity to shape both the
aperture and the search region.

\section{Examples}

A one-dimensional, unapodized aperture of width D has inverse Fourier
transform, $a(x)$, a Sinc function with its first zero at $x_0 = \lambda/D$.
We call $x_0$ the diffraction-limited spot size or, more simply, a DL
element.  The magnitude squared of $a(x)$ is $p(x)$, the
one-dimensional point spread function (PSF).  A square,
two-dimensional aperture has $PSF = p(x)p(y)$.  On the left in Figure 1
is the logarithm of the PSF for such an unapodized aperture.  The
diagonal line in the image shows where we will sample the PSF for
comparison purposes.  In all of our examples we assume that the 
available light is broadband with a bandwidth which is 20\% of the 
central wavelength.  Thus, the deep nulls in a conventional Sinc function 
are blurred out when we combine the PSF's at various wavelengths.

\epsscale{.8}
\begin{figure}[!ht]
\plotone{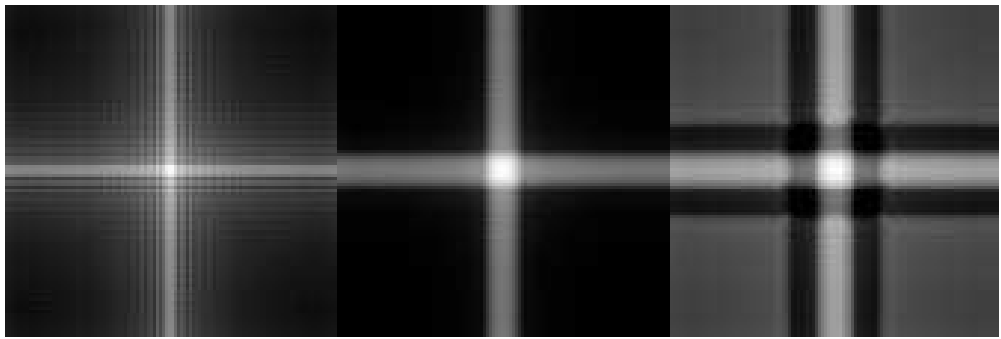}
\caption{Logarithm of Point Spread Functions for three apodizers.  
Left: No apodizer; Center: Product Prolate Spheroidal (PPS) with 
a $4.4 x_0$ by $4.4 x_0$ central square; Right: Modified PPS}
\end{figure}

 To form this first apodized aperture we space-limit a one-dimensional
 $f(x)$ to the interval, say, $-2.2 x_0 < x < 2.2 x_0$, then we band-limit
 the result to D.  (For these examples, we used FFT software with a
 calculation size of N = 256 and we set the aperture size to D = 41.)
 These space and bandwidth limiting operations are defined by
 Equations (3) and (4).  We repeat them until the iterations converge.
 What emerges is the prolate spheroidal function, $\phi(x)$.  Its
 Fourier transform, $\Phi(u)$, is the apodizer.  Following previous
 authors (i.e. Aime et al, 2002), we form the product aperture $P(u,v)
 = \Phi(u) \Phi(v)$, which we call a Product Prolate Spheroidal (PPS)
 aperture, and its inverse Fourier transform, p(x,y), is the
 telescope's PSF.  In the center of Figure 1 is the logarithm of the
 PSF for this apodizer.  Note that the PSF is concentrated along the x
 and y axes.  The aperture has a transmission of $24.9 \%$.  Papoulis
 (1968) showed that this aperture best concentrates the PSF into a central
 square of size $4.4 x_0$ by $4.4 x_0$.

Figure 1 shows a third PSF.  This was formed by defining a one-D
spatial mask, $w(x)$, which is 1 in the range from 0 to $2.27 x_0$,
zero from $2.27 x_0$ to $6 x_0$ and 1 for all larger x; and is
symmetric about $x = 0$.  The mask forces the PSF to be narrow, as
above, but also allows it to rise outside a restricted search region
from $2.27 x_0$ to $6 x_0$.  We apply the same iterative algorithm, starting
with the PPS apodizer and stopping the iterations when the
transmission is ($26.3 \%$). The result is the PSF shown on the right
of Figure 1.

Figure 2 shows a $45\deg$ cut through three PSF's.  The horizontal
axis plots diagonal distance in units of $x_0$ and the vertical axis plots
the logarithm of $p(x,y)$, scaled such that $p(0,0) = 1$.  Figure 2
also shows the PSF for the unapodized aperture.  Note that the PPS
plot is about seven orders of magnitude below the no-apodizer plot,
beyond about $3.5 x_0$. The modified PPS plot matches the PPS plot
from 0 to $3.5 x_0$, drops about 2 db below it to about $8 x_0$, then rises
almost to the no-apodizer plot.

\begin{figure}[!ht]
\plotone{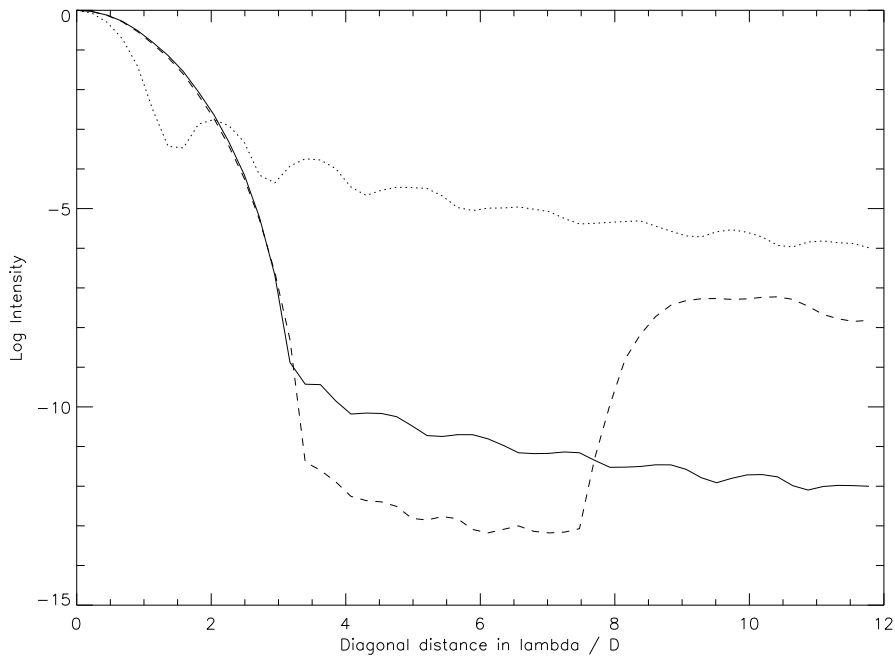}
\caption{Slices of the PSF's vs. distance in
Diffraction-Limited elements: No apodizer (dotted), Product Prolate
Spheroidal Apodizer (solid), Modified PPS Apodizer (dashed) }
\end{figure}

\begin{figure}[!ht]
\plotone{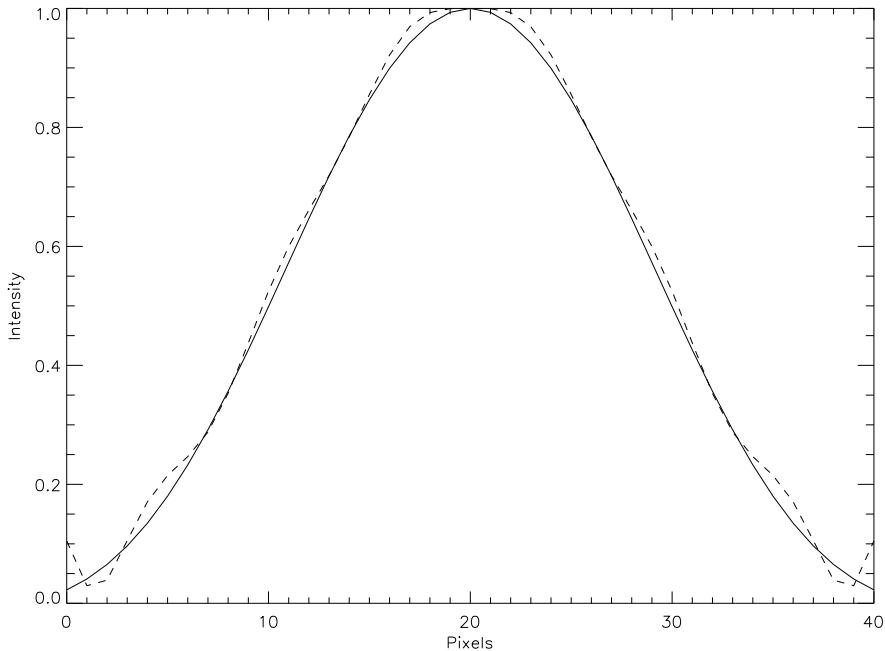}
\caption{Prolate Spheroidal Apodizer (solid), Modified PS Apodizer (dashed)}
\end{figure}

Figure 3 shows horizontal cuts through the PPS apodizer and the modified
PPS apodizer.

For our next set of examples we used a circular aperture of diameter
$D = 40$.  First we calculated the PSF for an unapodized aperture.  It
is the well-known Airy disc.  The logarithm of this PSF is shown on
the left of Figure 4.  (As in Figure 1 the PSF's are for wideband
light so the rings are somewhat indistinct.  The horizontal and
vertical bands at the edge of the figures are aliasing, caused by our
finite sampling.)

Next we formed a circular mask of radius 3.5z and exercised a
two-dimensional version of the iterative algorithm. This results in
the 2-D PSF in the center of Figure 4.  This is the PSF which Papoulis
(1968) reports but does not sketch.  We call the apodizer a Circular
Bessel (CB) apodizer.  Then we allowed the mask to rise at a radius of
$8.2 x_0$, so that the mask looks like a bulls eye, and exercised the two-D
algorithm.  The PSF is on the right of Figure 4.  Cuts through the
PSF's are shown in Figure 5.  Inside the search region the bulls eye
PSF is about two orders of magnitude below the PSF for the CB
aperture.  The apodizers are shown in Figure 6.  Both have
transmission of about $21\%$ .

\begin{figure}[!ht]
\plotone{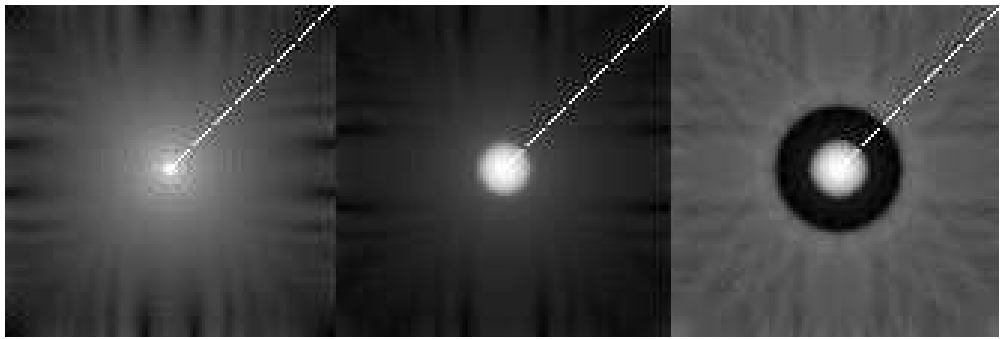}
\caption{Log PSF's for no apodizer, CB apodizer, and
Modified CB apodizer}
\end{figure}

\begin{figure}[!ht]
\plotone{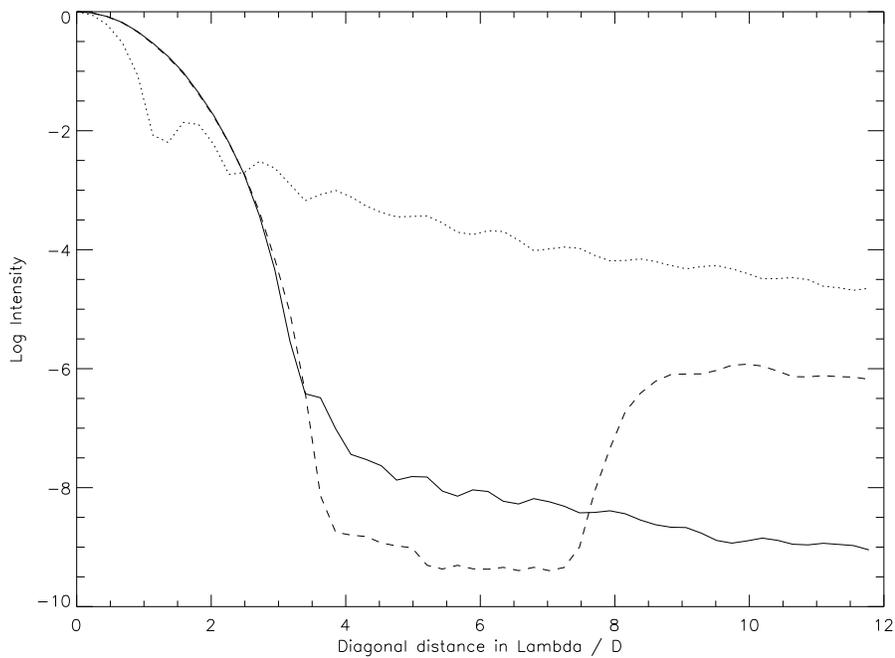}
\caption{PSF slices for no apodizer (dotted), CB (solid), and
Modified CB (dashed)}
\end{figure}

\begin{figure}[!ht]
\plotone{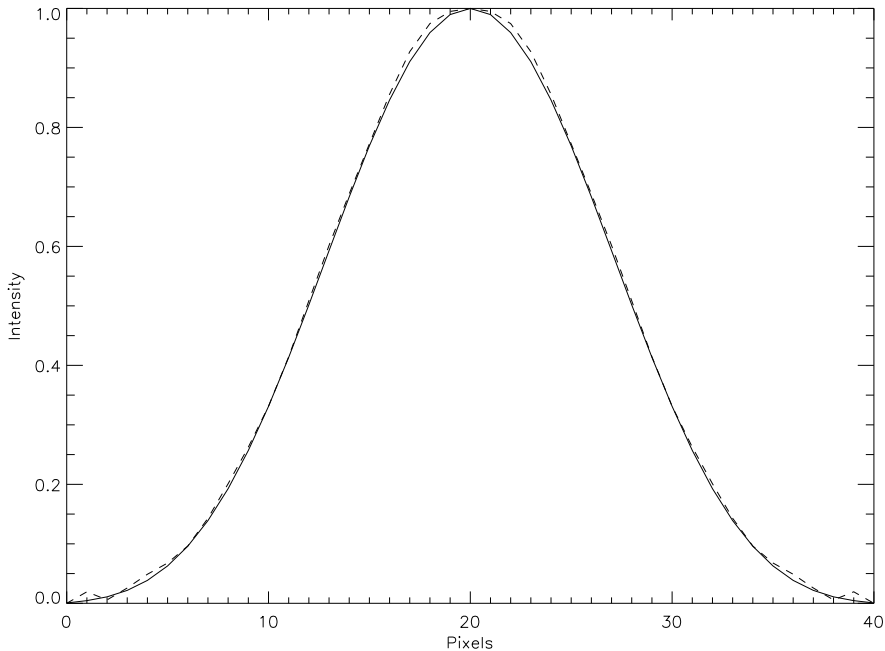}
\caption{CB apodizer (solid), Modified CB apodizer (dashed)}
\end{figure}

Finally, we formed an apodizer for a square aperture with a
two-dimensional mask which is a cross of width $4 x_0$, imbedded in a
circular mask of radius $8 x_0$, a hybrid apodizer.  It has a restricted
search area as shown on the left in Figure 7.  We exercised the
iterative algorithm, starting with the CB apodizer.  Figure 7 shows
the mask (the search area) and the log of the PSF.  Figure 8 shows a
slice through the PSF and, for comparison, a slice through the PSF for
the CB apodizer.  Despite what appear to be only subtle differences in
the shapes of the two apodizers, the hybrid apodizer has a
transmission of $39.3 \%$, almost double that of the CB apodizer; it
also has a narrower central peak and better sidelobe supression.  The
reduced search area of this hybrid apodizer provides significant
advantages over our earlier examples.  Figure 9 shows cuts through the
apodizer at 0 and at $45\deg$, which shows that the hybrid apodizer is
not circularly symmetric.

\begin{figure}[!ht]
\plotone{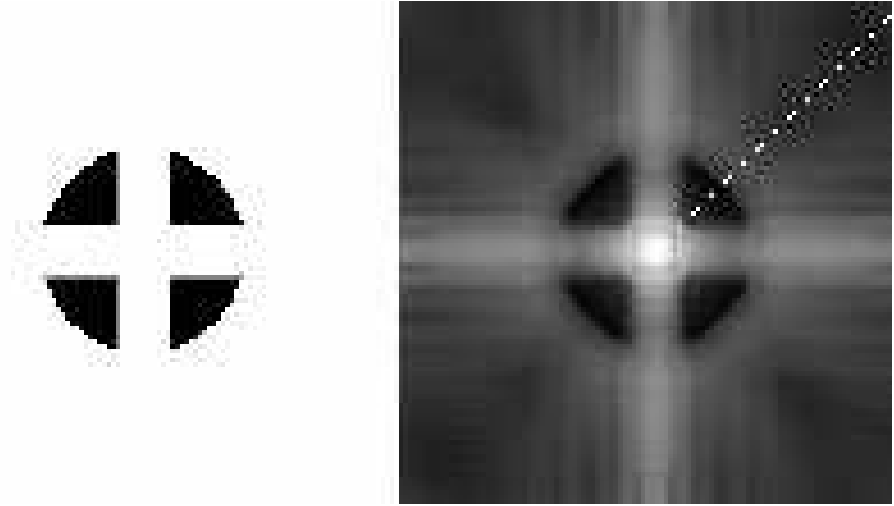}
\caption{Spatial mask and the log PSF for a hybrid apodizer,
	matched to the mask}
\end{figure}

\begin{figure}[!ht]
\plotone{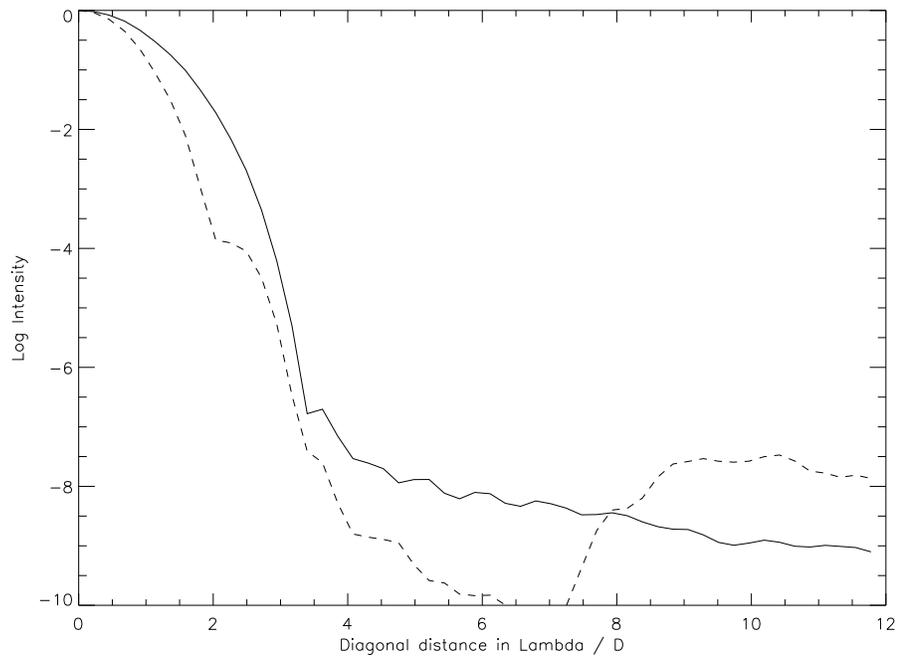}
\caption{PSF cuts: CB apodizer, $Trans = 21.7 \%$ (solid) ; Hybrid
 apodizer, $Trans = 39.3 \%$ (dashed)}
\end{figure}

\begin{figure}[!ht]
\plotone{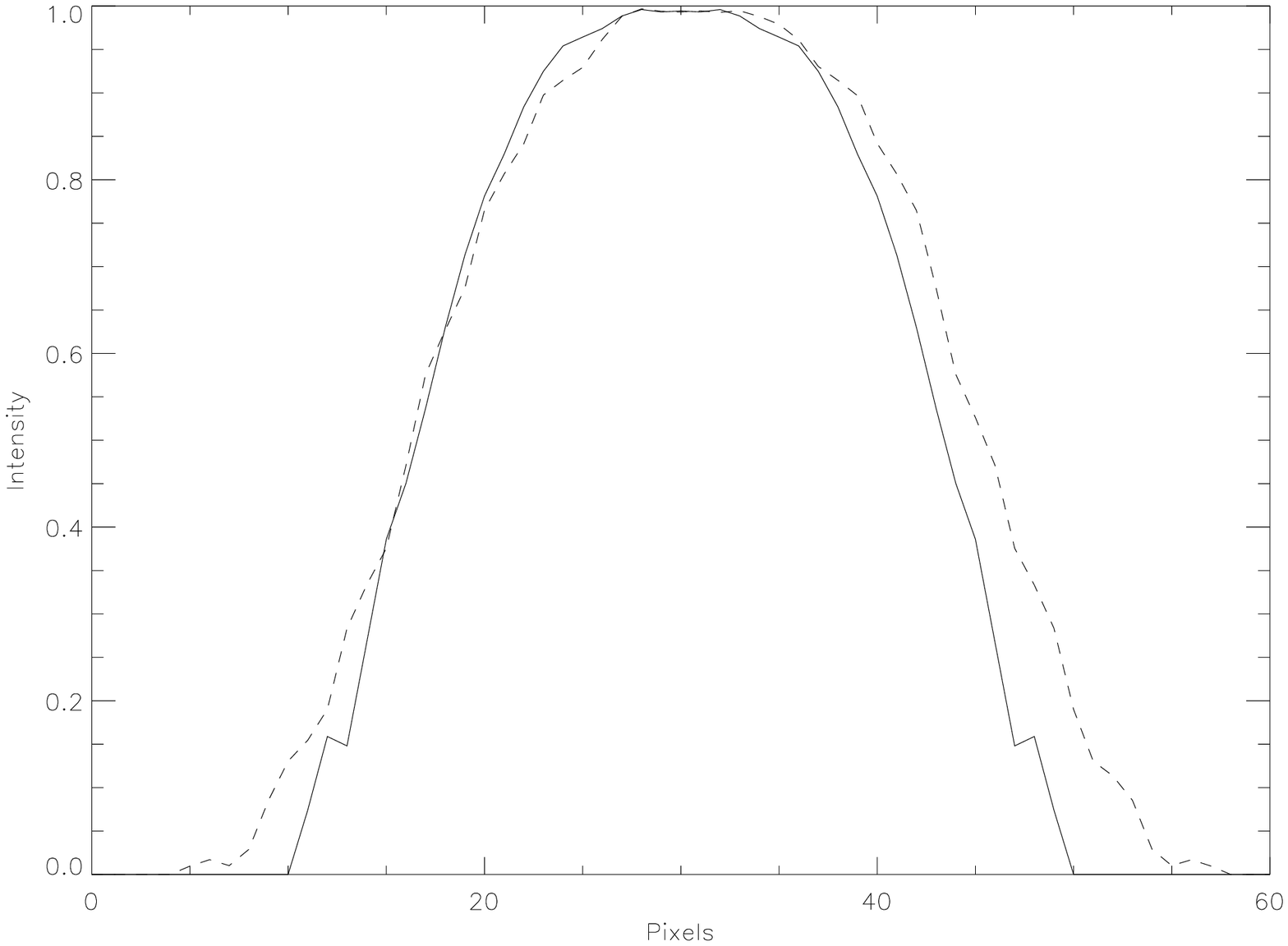}
\caption{Hybrid apodizer at $0\deg$ (solid) and at $45\deg$ (dashed)}
\end{figure}

\section{Discussion}

We have described an approach to designing an apodization mask that is
optimized for minimizing diffraction in a selected local area of the
focal plane. This would allow use of selectable apodization masks in a
TPF telescope that could, for example, reduce diffraction over a wide
area in the focal plane for a survey, and then produce deeper nulls in
local areas for spectroscopy of an already detected planet. Or one
could select an apodizer for the minimum width of the central peak,
trading off the size of the searchable region for minimum detectable
separation of star and planet. We also investigated use of apodizers
that vary only in one direction (constant in the other). These
apodizers result in maximizing throughput (close to 50\%) while
allowing detection of luminosity ratios of $10^8$. This is
insufficient for TPF-like missions (luminosity ratios will be of order
$10^{10}$ for an earth-like planet orbiting a star at 10 pc) but
hybrid apodizers, as described in the paper, allow a continuous
trade-off between maximizing throughput and minimizing diffraction.

The speckle due to optical surface errors and the diffraction will
produce a PSF that is highly centro-symmetric when (as required for a
TPF telescope) the phase errors are extremely small.  For very small
phases, the exponential phase can be accurately approximated by only
the linear term of the power series expansion of the exponential. So,
for example, a telescope with 1/1000 wave rms surface errors will have
a centro-symmetric psf to a precision of about 0.1\% and 1/10000 wave
results in centro-symmetry of 0.01\%. One can then use this asymmetry
to calibrate the residual speckle and diffraction in the half-plane of
the PSF that contains a planet using the opposite half-plane. 
Subtraction of the calibration region then increases the
dynamic range to an accuracy limited only by photon statistics.

Accurate intensity transmission of the apodizers is critical for 
their success - this is evident from the apparently small differences
in apodizer shape, as illustrated in Figure 3, producing substantial
effects on the PSF. We have estimated the precision required in the
transmission of the apodizer to be 0.3\% or better for it not to 
affect the dynamic range. This is the same requirement as one for
the uniformity of the reflective coatings of the telescope mirrors.
Numerical simulations that substantiate this result are included in
Nisenson et al. (2003)

The usual problems with iterative algorithms are choosing an initial
function and choosing a stopping point in the iterations.  For the
examples which yielded the PS and CB solutions, we started with a
Gaussian shape and we stopped the iterations when there was no
significant change in the solution, typically a few hundred
iterations.  For the other examples we started with a PS or CB
aperture. Continued iterations result in deeper and deeper suppression of
diffraction in the chosen region but also resulted in decreasing 
throughput for the apodizer. We stop the iterations when the 
diffraction is reduced to a selected level while leaving the throughput at
a maximum.

We have shown an approach to calculating an optimum apodizer that
maximizes the detectable luminosity ratio for exo-planet detection in
the focal plane of the TPF telescope.  The next step in this analysis
would be to include realistic specifications on wavefront errors from
the telescope optics and also to factor in practical limits on the
accuracy in the manufacture of apodizing masks. How well continuous
tone apodizers can be manufactured is an open question.  Many of the
techniques developed for photolithography allow very precise shaped
transmission to be generated in special glass. Such a mask was used in
a laboratory demonstration of imaging two close point sources with a
luminosity ratio of nearly $10^9$ (Melnick et al, 2001).

\end{document}